\documentclass[12pt,preprint]{aastex}

\newcommand{\grb}{GRB~$980425$}

\newcommand{\cxo}{{\it CXO}}
\newcommand{\hst}{{\it HST}}
\newcommand{\batse}{{\it BATSE~}}
\newcommand{\sax}{{\it BeppoSAX}}
\newcommand{\xmm}{{\it XMM}}
\newcommand{\nfi}{{\it NFI}}

\newcommand{\sn}{{SN~1998bw}}

\newcommand{\host}{{\it ESO~$184-$G82}}

\newcommand{\etal}{{\it et al.~}}

\newcommand{\lSect}[1]{{\label{sec:#1}}}

\newcommand{\Sectff}[1]{{\ref{sec:#1}}}
\newcommand{\Sect}[1]{{\S~\Sectff{#1}}}

\def\ltaprx {\lower .1ex\hbox{\rlap{\raise .6ex\hbox{\hskip .3ex
        {\ifmmode{\scriptscriptstyle <}\else 
                {$\scriptscriptstyle <$}\fi}}}
        \kern -.4ex{\ifmmode{\scriptscriptstyle \sim}\else 
                {$\scriptscriptstyle\sim$}\fi}}}
\def\gtaprx {\lower .1ex\hbox{\rlap{\raise .6ex\hbox{\hskip .3ex
        {\ifmmode{\scriptscriptstyle >}\else 
                {$\scriptscriptstyle >$}\fi}}}
        \kern -.4ex{\ifmmode{\scriptscriptstyle \sim}\else 
                {$\scriptscriptstyle\sim$}\fi}}}

\slugcomment{Submitted to ApJ, \today} 

\shorttitle{X-Ray Observations of SN 1998bw}
\shortauthors{Kouveliotou et al.}

\begin{document}

\title{Chandra Observations of the X-ray Environs of
SN~1998bw/GRB~980425}

\author{C.~Kouveliotou\altaffilmark{1,2}, 
S.~E.~Woosley\altaffilmark{3},
S.~K.~Patel\altaffilmark{2},
A.~Levan\altaffilmark{4},
R.~Blandford\altaffilmark{5}, 
E. Ramirez-Ruiz\altaffilmark{6},
R.A.M.J.~Wijers\altaffilmark{7}, 
M.~C. Weisskopf\altaffilmark{1},
A.~Tennant\altaffilmark{1},
E.~Pian\altaffilmark{8},
P.~Giommi\altaffilmark{9}}

\email{chryssa.kouveliotou@nasa.gov}

\altaffiltext{1} {NASA/Marshall Space Flight Center, NSSTC, SD-50, 320
Sparkman Dr., Huntsville, AL 35805, USA}

\altaffiltext{2} {Universities Space Research Association, NSSTC,
SD-50, 320 Sparkman Dr., Huntsville, AL 35805, USA}

\altaffiltext{3} {Department of Astronomy \& Astrophysics, University
of California at Santa Cruz, Santa Cruz, CA 95064, USA}

\altaffiltext{4} {Department of Physics and Astronomy, University of
Leicester,University Road, Leicester, LE1 7RH, UK}

\altaffiltext{5} {Kavli Institute for Particle Astrophysics and
Cosmology, Stanford, CA 94305, USA}

\altaffiltext{6} {Institute for Advanced Study, Olden Lane, Princeton,
NJ 08540, USA; Chandra Fellow}

\altaffiltext{7} {Astronomical Institute ``Anton Pannekoek''and Center
for High Energy Astrophysics, University of Amsterdam, Kruislaan 403,
1098 SJ Amsterdam, NL}

\altaffiltext{8} {INAF, Osservatorio Astronomico di Trieste, Via
G.B. Tiepolo 11, I-34131 Trieste, I}

\altaffiltext{9} {ASI Science Data Center, c/o ESRIN, Via G. Galilei,
I-00044 Frascati, I}

\begin{abstract}
We report X-ray studies of the environs of \sn\ and \grb\ using the
Chandra X-Ray Observatory 1281 days after the GRB. Eight X-ray point
sources were localized, three and five each in the original error
boxes - S1 and S2 - assigned for variable X-ray counterparts to the
GRB by \sax. The sum of the discrete X-ray sources plus continuous
emission in S2 observed by \cxo \ on day 1281 is within a factor of
1.5 of the maximum and the upper limits seen by \sax. We conclude that S2 is the
sum of several variable sources that have not disappeared, and
therefore is not associated with the GRB. Within S1, clear evidence is
seen for a decline of approximately a factor of 12 between day 200 and
day 1281.  One of the sources in S1, S1a, is coincident with the
well-determined radio location of \sn, and is certainly the remnant of
that explosion. The nature of the other sources is also
discussed. Combining our observation of the supernova with others of
the GRB afterglow, a smooth X-ray light curve, spanning $\sim1300$
days, is obtained by assuming the burst and supernova were coincident
at 35.6 Mpc. When this X-ray light curve is compared with those of the
X-ray ``afterglows'' of ordinary GRBs, X-ray Flashes, and ordinary
supernovae, evidence emerges for at least two classes of lightcurves,
perhaps bounding a continuum. By three to ten years, all these
phenomena seem to converge on a common X-ray luminosity, possibly
indicative of the supernova underlying them all. This convergence
strengthens the conclusion that \sn\ and \grb\ took place in the same
object. One possible explanation for the two classes is a (nearly)
standard GRB observed at different angles, in which case X-ray
afterglows with intermediate luminosities should eventually be
discovered. Finally, we comment on the contribution of GRBs to the ULX
source population.
\end{abstract}


\section{Introduction}
\lSect{intro}

One of the most exciting developments in the study of gamma-ray bursts
(GRBs) was the discovery, in 1998, of a GRB apparently in coincidence
with a very unusual supernova of Type Ic (Galama \etal\ 1998). This
coincidence of \sn\ and \grb\ offered compelling evidence that GRBs
are indeed associated with the deaths of massive stars, and that, at
least in some cases, GRBs go hand in hand with stellar explosions
(Woosley 1993; MacFadyen \& Woosley 1999). The large energy release
inferred for the supernova also suggested a novel class of explosions,
called by some ``hypernovae'' (Paczy\'nski 1998), having unusual
properties of energy, asymmetry, and relativistic ejecta.

However, this identification was challenged on two grounds. First,
there were {\sl two} variable X-ray sources identified with \sax\ in
the initial 8.0$^{\prime}$ radius GRB error box; one was not the
supernova. Second, if it were associated with the nearby supernova,
\grb\ was a most unusual burst with gamma-ray energy per solid angle
roughly four orders of magnitude less than typical. Further, the \sax\ decay of the \sn\ associated X-ray source was much slower than the typical GRB X-ray afterglow decay (Pian \etal\ 2000). Interestingly, observations (Pian \etal\ 2003) of the two sources in March 2002 with the X-ray Multi-Mirror (\xmm) telescope revealed that the X-ray emission of the \sn\ associated \sax\ error box had decreased at a faster pace than expected by a simple extrapolation of the earlier measurements. Moreover, the second \xmm\ source was found to consist of a number of faint point-like sources, whose integrated emission was consistent with the average brightness measured with \sax\ (Pian \etal\ 2000).

Strong support for the GRB-SN association came from the very recent spectroscopic detection (Hjorth \etal 2003; Stanek \etal\ 2003) of a supernova (SN~2003dh) in the optical light curve of GRB~030329. In this case, the SN detection was obscured by the extreme optical brightness of the GRB afterglow. While the models of MacFadyen and Woosley (1999) suggest that there can be adequate $^{56}$Ni to make the GRB-supernova bright, it is now becoming evident that we must also account for the potentially much brighter optical afterglow of the GRB itself. 

Here we report the results of a study of the environs of \sn\ and
\grb\ using the Chandra X-Ray Observatory (\cxo) 1281 days after the
GRB. This study had several goals. First, given the intervening three
years, has the evidence strengthened for the GRB-SN association?  We
believe that it has (\Sect{xlite}).  Second, how does the X-ray light
curve of GRB 980425, measured across 1300 days, compare
with those of other GRBs and with other kinds of high energy
transients - in particular X-ray flashes (XRFs; Heise 2003) and
supernovae?  What does the comparison tell us about the nature and
origin of GRBs?  We find that it provides evidence for a common theme
underlying all these events - a powerful asymmetric supernova with
relativistic ejecta along its polar axes and an observable event that
varies depending upon the viewer's polar angle (\Sect{systematic} and
\Sect{theory}).

Finally, we are interested in the environs of SN 1998bw.  Aside from
the one supernova, does this region show evidence for unusual stellar
activity as might characterize a vigorously star forming region
(\Sect{sources})?  \sn\ offers the best opportunity to study a GRB
site up close, and one should take every advantage of that.

\section{Prompt Observations of \grb\ and \sn} 
\lSect{discov}

\grb\ triggered the Burst And Transient Source Experiment (\batse) on
board NASA's Compton Gamma-Ray Observatory ({\it CGRO}) on 1998 April
25, 21:49:09 UT; the event was simultaneously detected by the \sax\
Gamma-ray Burst Monitor ({\it GRBM}) and Wide Field Camera ({\it
WFC}). The burst consisted of a single peak of $\sim23$ s duration,
with peak flux and fluence (24-1820 keV) of $(3.0\pm0.3)\times10^{-7}$
erg/cm$^2$ s, and $(4.4\pm0.4)\times10^{-6}$ erg/cm$^2$,
respectively. Galama \etal (1998) observed the {\it WFC}
8.0$^{\prime}$ error box with the New Technology Telescope (NTT) at
the European Southern Observatory (ESO) on April 28.4 and May 1.3 UT
and, in the error box of \grb, they found supernova \sn, located in an
HII region in a spiral arm of the face-on barred spiral galaxy
ESO~$184-$G$82$, at z=0.0085, corresponding to a distance of 38.5 Mpc
(Galama et al. assumed a Hubble constant, $H_0=65$ km/s Mpc, but in
the following we use $H_0=72\pm8.0$ km/s Mpc as measured by Freedman
et al. (2003), placing \sn\ at 35.6 Mpc).

On 1998 April 26-28, ten hours after the GRB, the \sax\ Narrow Field
Instruments (\nfi) observed (Pian \etal\ 2000) the {\it WFC} error box
and revealed two previously unknown, weak X-ray sources,
1SAX~J$1935.0-5248$ (hereafter S1) and 1SAX~J$1935.3-5252$ (hereafter
S2) with an uncertainty radius of 1.5$^{\prime}$ each. S1 included the
\sn, but S2 was $4.5^{\prime}$ away. Both sources were observed two
more times with the {\it NFI} resulting, in the case of S2, in two
detections and two upper limits. In contrast, during the six month
interval spanned by all \nfi\ observations, the flux, $F$, of S1
followed a power-law temporal decay (Pian \etal 2000): $ F_{2-10 keV}
= 4.3(\pm0.5)\times10^{-13} (t/{\rm 1day})^{-0.2}$ erg/cm$^2$ s, a
much flatter trend than the one observed for other GRB X-ray
afterglows, but a decaying trend nevertheless.

At 35.6 Mpc the apparent isotropic energy of \grb\ ($7 \times 10^{47}$ erg) was
about four orders of magnitude smaller than that of `normal' GRBs
(Bloom \etal 2003). Moreover, independent of its connection with a
GRB, \sn\ was extraordinary in many ways. Its light curve resembled a
Type Ia supernova in brilliance, but the spectrum was more like Type
Ic (H, He, and Si lines absent, but the spectrum was peculiar even for
Ic). Thus, the two phenomena together presented a very interesting
scientific puzzle whose solution required the combined superb
resolution of the Hubble Space Telescope (\hst) and \cxo.

\section{\cxo\ Observations at Day 1281}
\lSect{cxoobs}

\setcounter{footnote}{0}

\cxo\ observed S1 and S2 on 2001 October 27, for a total time on source
of 47.7~ks. S1 fell completely on ACIS-S3 (a back-illuminated CCD) and
S2 only partially, with most of the error region falling on ACIS-S2 (a
front-illuminated CCD), both operating in Time Exposure (TE) mode. The
data were processed using the CIAO
(v3.0.1)\footnote{http://asc.harvard.edu/ciao/} and
CALDB(v2.34)\footnote{http://asc.harvard.edu/caldb/} software. More
specifically, we used the CIAO tool {\it acis-process-events} to
ensure that the latest gain corrections were applied (those
corresponding to our observation date). Further, we removed the
standard pixel randomization, applied CTI corrections, filtered the
data to include events with ASCA grades = 0,2,3,4, and 6, and applied
standard GTIs. We corrected the systematic offset in the aspect using
the {\it fix-offset}
thread\footnote{http://cxc.harvard.edu/cal/ASPECT}; this results in an
offset of $\delta{\rm RA} = 0.16\arcsec$ and $\delta{\rm DEC}=
0.43\arcsec$. To improve the ACIS-S3 spatial resolution we use the
method described by Mori \etal (2001) and Tsunemi \etal (2001) to
adjust the event locations. These studies conclude that by defining
the event location on a pixel as a function of the event grade (rather
than placing the event at the pixel center) one can achieve $\sim10\%$
improvements in spatial resolution. During our observation the X-ray
background increased by $\sim50\%$ with variations on time scales of a
few kiloseconds. Since the point source emission is not significantly
masked by such slow background variations, we chose to include all the
data in our analysis to preserve our (limited) source counts.

\subsection{S1 and S2: Source Identification, Locations and Energetics}
\lSect{sources}

We used the source-finding method originally described in Swartz
\etal (2002), accepting as detections all sources with a minimum
signal-to-noise ratio (S/N) of 2.6. For source detection purposes we
searched images consisting of data between $0.3-8.0$ keV to avoid the
ACIS high energy background. We discuss below the sources within the 
$1.5\arcmin$ radius ($1\sigma$) \nfi\ error circles of S1 and S2 only.

\subsubsection{S2}

S2 was resolved into five sources (S2abcde; Figure 1c and Table
1). Given the very limited count number per source, we combined four
of them in one spectral fit with a power law function and a hydrogen
column density, $N_{\rm H}= 3.95\times10^{20}$ cm$^{-2}$ (corresponing to the Galactic absorption in the line of sight to \grb; Schlegel, Finkbeiner \& Davis 1998). Here we
assume a spectral similarity between these sources to mitigate the
difficulty of fitting individual spectra of very few counts each. This
order of magnitude approximation is acceptable within our source
statistics (see also Figure 2, where rudimentary count spectra are
presented for S1abc, and S2ce). However, one source (S2d), fell into
the gap between the ACIS CCDs S2 and S3 and its total counts had to be
adjusted taking into account the reduced gap exposure time.  Using the
spectral index of -1 from the fit, we applied a conversion factor of
$1.89\times10^{-11}$ erg/cm$^2$ s ($0.3-10.0$ keV) per count/sec
($0.3-8.0$ keV) to the S2d counts and finally estimated the total
(combined) flux in S2 ($0.3-10$ keV) to be $3.0(3)\times10^{-14}$
erg/cm$^2$ s (here and in Table 1 the numbers in parentheses
correspond to the $1\sigma$ errors in the last digit). 

The \sax\ flux for S2 was, however, calculated for a much larger extraction radius (3$\arcmin$) to account for the extreme faintness of the detection within a 1.5$\arcmin$ radius, which was at the level of the \nfi\ confusion limit (Pian \etal\ 2000). Only $\sim60 \%$ of this larger error circle is covered with \cxo. Using the same source-finding algorithim criterion described above, we found a total of 18 sources within the enlarged area, with a total flux of $S2(3\arcmin)=1.6\times10^{-13}$ erg cm$^{-2}$ s. This is roughly the flux to be compared to the \sax\ S2 value at day 1. Subsequently we re-calculated the
total \sax\ flux within S2 at day 1, using the same spectral function
derived with \cxo\ and found it to be $4\times10^{-13}$ erg/cm$^2$
s (with $<2\sigma$ significance). Taking into account the partial coverage with \cxo\ of the \nfi\ error circle of S2, we estimate that the day 1281 \cxo\ flux value is within a factor of 1.5 of the $3\sigma$ detection limit of \sax, indicating that at best there was no siginificant variation of the sources within S2 over the last 3.5 years. This result is consistent with the \xmm\ observations of S2 (Pian \etal\ 2003). We further discuss the
evolution of the light curve of S2 in \Sect{xlite}.

\subsubsection{S1}

We initially identified two sources $36\arcsec$ apart within the S1
(1.5$\arcmin$) error region\footnote{Here we are only considering the area corresponding to the \nfi\ half power radius of $1.5\arcmin$, as most of the \sax\ signal for S1 at day 1 was within this area (Pian \etal\ 2000). An enlarged radius region (3$\arcmin$) for S1 results in a total flux of $2.7\times10^{-13}$ erg cm$^{-2}$ s, almost half the \sax\ value at day 1}, one of which coincided with the location of \sn. However, further inspection of the latter resolved this source into two with a radial separation of $\sim1.5\arcsec$ (Figure 1b). We
fitted these sources simultaneously with two 2-D circular Gaussians to
better estimate their centroids and found that they are both
consistent with point sources. Hereafter we designate the X-ray
sources detected within the S1 error region as S1a (corresponding to
\sn), S1b, and S1c. We have fitted a power law to each unbinned
non-background subtracted source spectrum, assuming the same $N_{\rm
H}$ as for S2. Our fit parameters are derived using the C-statistic
(Cash 1979), appropriate for low count data (Table 1). 

Further, we reanalyzed all archival \hst\ and {\it ESO/Very Large
Telescope (VLT)} observations of ESO~$184-$G82, the host galaxy of
\sn, concentrating on the immediate environment of the supernova. The
small field of view of \hst/{\it STIS} ($\sim$ 50 \arcsec) contains
only the sources S1a and S1b. In order to obtain accurate astrometry
we, therefore, registered both the \cxo\ and \hst\ images
independently to an R-band observation obtained at the {\it VLT} on
1999 April 18. The $3.4\arcmin \times 3.4\arcmin$ field of the {\it
VLT} image contained three additional X-ray sources with apparent
optical counterparts which we used to align the two fields. Finally,
we aligned the \hst\ and {\it VLT} images using eight, non-saturated
point sources present in each image (both alignments were performed
using {\it IRAF} and the tasks {\sc geomap \& geoxytran}). We were
then able to project the relative position of S1a and S1b onto the
\hst\ field with a positional accuracy of $\sim$ 0\farcs3. This is
shown schematically in Figures 1e,f, where we zoom in the region
around S1a and S1b for both the \cxo\ and \hst\ data. It is obvious in
the \hst\ image that while there is a clearly identified optical
counterpart to the \sn, there is no variable counterpart within the
$\sim$ 0\farcs3 \cxo\ error circle ($1\sigma$ radius) of S1b; we do
however, see a variable source just outside the error region. Levan
\etal\ (in preparation) present a detailed study of this source
together with three more transients within a radius of $6\arcsec$ of
the \sn\ as well as narrow field spectroscopy of the \sn\
environment. A counterpart search for the other six sources in S1 and
S2 in all available catalog data\footnote{http://heasarc.gsfc.nasa.gov/db-perl/W3Browse/w3browse.pl} failed to identify any known objects at their positions.

     \begin{center}
      \begin{tabular}{rcccccccc}
      \multicolumn{7}{c}{{\bf Table 1:} \cxo\ Sources Within S1 and 
S2}\\

      \hline
      \hline
      ID & RA & DEC & Counts$^a$ & S/N  & Index$^b$ & Flux$^c$  & Luminosity$^{c,d}$  \\
      & (hms) & ($\arcdeg~\arcmin~\arcsec$) & & & & 
erg cm$^{-2}$ s$^{-1}$ & erg s$^{-1}$ \\

      \hline
%% S1 %%
      S1a  & 19 35  3.31 & -52 50  44.8 & 24 & 4.53 & 1.0(3) & $8(2)  
\times10^{-15}$ & $1.2\times10^{39}$   \\
      S1b  & 19 35  3.23 & -52 50  43.4 & 52 & 6.47 & 1.6(2) & 
$9(1)\times10^{-15}$ & $1.4\times10^{39}$   \\
      S1c  & 19 35  0.56 & -52 50  21.2 & 27 & 4.69 & 0.5(3) & 
$1.2(2)\times10^{-14}$ & --                   \\
\hline
%% S2 %%
      S2a$^e$  & 19 35 25.75 & -52 54  19.0 & 10 & 2.75 & --     & 
$4(1)\times10^{-15}$ & --                   \\ 
      S2b$^e$  & 19 35 25.22 & -52 54  21.2 &  7 & 2.66 & --     & 
$3(1)\times10^{-15}$ & --                   \\ 
      S2c     & 19 35 24.65 & -52 54  40.7 & 22 & 4.46  & 1.7(4) & 
$5(1)\times10^{-15}$ & --                   \\ 
      S2d$^f$ & 19 35 23.15 & -52 53   5.6 &  9 & 2.82 &  --  & 
$8(3)\times10^{-15}$ & --                   \\
      S2e     & 19 35 14.03 & -52 53  51.9 & 23 & 4.19 & 0.7(3)  & 
$1.0(2)\times10^{-14}$  & --                   \\ 
      \hline
%     \multicolumn{7}{l}{
\end{tabular}
\end{center}
{\footnotesize{
$^a$between $0.3-8.0$ keV, $^b$for an energy spectrum of $\propto E^{-\gamma}$, $^c$between $0.3-10.0$ keV,
$^d$assuming the source is in \host\ at 35.6 Mpc, 
$^e$fluxes for sources with less than 20 counts are derived using a PL
index of -1.0, and a conversion factor of $1.89\times10^{-11}$ ergs
s$^{-1}$ cm$^{-2}$ ($0.3-10.0$ keV) per 1 count/s ($0.3-8.0$ keV),
$^f$source flux corrected for decreased exposure}}

From Table 1, we notice that the X-ray luminosities of S1a and S1b
marginally exceed the Ultra Luminous X-ray source (ULX) threshold
luminosity of $L=1\times10^{39}$ erg/s (Fabbiano 1989).\footnote{Here we are
not discussing the nature of S1c as there is no clear evidence for the
association of this source with ESO~$184-$G82} Further, inspection of
the S1b light curve in Figure 2 reveals that the source was `on' during
the first 20 ks of our \cxo\ observation, remained `off' for the
following $\sim28$ ks and possibly turned on again during the last 2
ks of the observation. Counting the GRB-SN source also as a ULX, we then have two of these sources in \host. What is the probability that S1b resides in \host, how common are ULXs in galaxies and what is the
nature of S1b, this ultraluminous X-ray transient?

To address the first question we use the log $N$-log $S$ relation for the hard X-ray sources by Giacconi \etal\ (2001): $N(>S) = 1200\times($ $S \over 2\times10^{-15}$ $)^{-1\pm0.2}$ sources/deg$^2$ (the one for the soft sources is smaller, so we use this as a conservative upper limit). Substituting for $S=9\times10^{-15}$ erg/cm$^2$ s, we find $N(>S)=400$ sources/deg$^2$. Correspondingly, we expect 0.01 source within a radius of 10\arcsec. We conclude that S1b is most likely associated with the spiral arm of \host\ and is not a background source. 

There are two extensive studies of ULX populations in normal galaxies
(Roberts and Warwick 2000; Humphrey \etal\ 2003) based on data from
{\it Rosat} and \cxo, respectively. While both studies find
substantial evidence for a correlation between the ULX count number
with star forming mass, the {\it Rosat} survey tends to underestimate
their count numbers by a factor of $5-10$ with respect to the \cxo. We have, therefore, used
the {\it Rosat} luminosity distribution
DN/dL$_{38}=(1.0\pm0.2)$L$^{-1.8}_{38}$, normalized to a
$B$-band luminosity of $10^{10}$L$_{\odot}$ to calculate the expected
number of ULXs in ESO~$184-$G82, multiplied by a factor of 10 as
indicated by the \cxo\ survey. We derive a total corrected expected
number of 0.28 ULXs in \host, a factor of $\sim7$ less than the actual
observed number of 2. We conclude that if S1a and S1b are both in \host, there is
a somewhat unusual concentration of ULXs in this galaxy. Finally, \host\ is one of 5
members of a galaxy group. In the \cxo\ data, which covers all
members, we do not detect any other X-ray sources from this galaxy
cluster. This practically means that, assuming that all members are at
the distance of \host, we detect only 2 ULXs from the whole cluster (one of which is a GRB/SN afterglow), both within a radius of $6\arcsec$. We discuss the results of narrow
band photometry of \host\ together with the properties of other
unusual transients found in the \hst\ data in the \sn\ environment in
detail in Levan \etal\ (in preparation).

To test S1b for variability we extracted the source lightcurve
in 1000 s bins and performed a Kolmogorov-Smirnov test against 
a constant flux distribution; this is the prefered method for the low
number of counts ($<100$) which are seen in the source. We derive $P_{KS}=0.05$
(where $P_{KS}$ is the probability that the source is constant); thus S1b is consistent with being variable at the 95\% confidence level. The transient nature and the energetics of S1b indicate a possible microquasar similar to e.g., GRO~J1915$+$105 in our own Galaxy
(Mirabel \& Rodriguez 1994; Greiner, Morgan \& Remillard 1996). This
superluminal source exhibits transient behavior in a wide variety of
time scales and intensities, and has an apparent (isotropic) X-ray
luminosity of $\sim7\times10^{39}$ erg s$^{-1}$, well above its
Eddington limit (Greiner, Cuby \& McCaughrean 2001). Fabbiano \etal\
(2003) have found in a study with \cxo\ of the ULX sources in the
Antennae galaxies that seven out of the nine ULXs are time variable,
most likely accreting compact X-ray binaries. Interestingly, the
average co-added spectrum of the Antennae ULXs resembles that of
Galactic microquasars (Zezas \etal\ 2002). We conclude that most
likely S1b is a microquasar in \host.

\section{Systematics of the X-ray light curves of High Energy Transients}
\lSect{systematic}

Figure 3 shows, on a common scale, all GRB afterglows with
measurements covering several tens of days. There are, unfortunately,
only a few curves because such measurements can only be made on GRBs
that are relatively nearby, but their light curves should be
illustrative.  For GRB~$030329$, we used data points from both the
Rossi X-Ray Timing Explorer (RXTE) and \xmm\ reported by Tiengo \etal\ (2003). The original four points of \grb\ were reported by Pian \etal\ (2000); the fifth point is the
\cxo\ observation of this paper. For GRBs 021004, 010222, 000926, and
970228, we used data from Sako \& Harrison (2002ab), Bj{\"
o}rnsson \etal\ (2002), Harrison \etal\ (2001), and Costa \etal\
(1997), respectively. We reanalyzed all data, calculated the fluxes in
the same energy interval ($0.3-10$ keV) and then converted them into
(equivalent isotropic) luminosities using a cosmology of $\Omega_{\rm
M}=0.27, \Omega_{\Lambda}=0.73$, and $H_0=72$ km s$^{-1}$ Mpc$^{-1}$
to eliminate the distance dependence.  As has been noted many times,
\grb\ and its early afterglow falls orders of magnitude below the
``ordinary'' GRBs.

We then compared these curves with both those of supernovae of all
types and with XRFs. We collected, reanalyzed and converted flux data
to luminosities as described above for the Type IIb SN~1993J (Uno
\etal\ 2002; Swartz \etal\ 2002), the Type Ic supernovae 1994I and 2002ap
(Immler \etal\ 2002; Sutaria \etal\ 2003) and the Type II supernovae 1998S
and 1999em (Pooley \etal\ 2002). Each source is indicated with a
different symbol in Figure 3.  Finally we added the three XRF
afterglow light curves available (011030, 020427, and 030723; Bloom
\etal\ 2003; Butler \etal\ 2003) assuming a redshift of $z=1$ for each
of them (there are no redshift measurements for any of these sources). The slopes of the XRF afterglows agree well with
those of the typical GRBs and their luminosities are comparable for
the distances assumed. Recently, Soderberg \etal (2003b) have reported
the counterpart identification of XRF~020903 at a redshift
$z=0.251$. To reflect this lower distance scale for XRFs, we plot on
Figure 3 another set of light curves (dashed lines) corresponding to
the luminosities all XRFs mentioned above would have, when placed at a
distance of $z=0.251$. They still fall well within the `typical' GRB
range; thus XRFs would have to be extremely nearby for their X-ray
light curves to be distinct from the generic GRB X-ray afterglow.

The resulting plot is striking in several ways. Despite the huge
disparity in initial appearance, there are compelling indications of a
common convergence of all classes of phenomena plotted - GRBs, \grb,
XRFs, and the most energetic supernovae - to a common resting place, L
$\sim 10^{39}-10^{40}$ erg s$^{-1}$ about three to ten years after the explosive
event. \grb, being the closest by far of any GRB ever studied, has the
virtue of being followed all the way to the ``burial ground'', but a
simple logarithmic extrapolation of the GRB and XRF light curves
places them squarely in this region as well.  As the collapsar model has
predicted (MacFadyen \& Woosley 1999) and observations of SN~2003dh/GRB~030329 have unambiguously confirmed for one case (Hjorth
\etal\ 2003; Stanek \etal\ 2003), an energetic supernova is expected
to underly all GRBs of the long soft variety (Kouveliotou \etal\
1993). Zhang, Woosley \& Heger (2003) have also predicted that a
similar supernova will underly all XRFs.

We proceed now to a wider GRB X-ray afterglow comparison with those
plotted on Figure 3. We have often observed GRB afterglows with
temporal decays which cannot be described by a single power
law. Rather, these afterglows initially decay as $t^{-1}$ or a bit
steeper, and at later times the decay index becomes approximately
$-2$. This steepening is attributed to the fact that the
ultrarelativistic outflow is initially collimated within some angle
$\theta$ (typically thought to be of order 10 degrees, Frail et~al.\
2001; see also Van Paradijs et~al.\ 2000, and references therein). The
transition to a steeper power law marks the time when the outflow
begins to expand laterally, which occurs around the time when
$\theta\sim1/\Gamma$, where $\Gamma$ is the Lorentz factor of the
blast wave.  Typically, the `break time' when this happens is around
$0.3-3$ days after the burst, but a few cases have much later breaks,
if any (e.g., GRB\,970508, 000418; Frail et~al.\ 2001).  If this is
also true for the X-ray afterglows shown in Fig.~3, then they should
be extrapolated to later times with a steeper power law than the
average of the data shown. In that case, they would reach the
luminosity level of SNe sooner, and after that evolve like the SNe,
because the blast wave will have become non-relativistic. Thereafter,
the system may simply evolve like an X-ray supernova with an energy
that has been augmented with that of the initially relativistic blast
wave. However, it is tantalizing that a few systems, e.g., GRB\,030329
do not yet show such a steep decay even at 30-40 days, possibly
indicating that other mechanisms than the standard collimated
afterglow contribute to the emission. For example, it has been
proposed that the outflows of GRBs are structured (e.g., M\'esz\'aros
et al.\ 1998, Rossi et~al.\ 2002, Ramirez-Ruiz et al. 2002), i.e.,
they eject material with ever lower Lorentz factors at ever larger
angles from the jet axis, and this slower material affects the
afterglow at ever later times, making it decay more slowly. In this
context, it is interesting to note that in GRB\,030329 there is
evidence for two jet breaks, one at 0.5 days and one at about one week
(Berger et al.  2003) from radio to X-ray data. Moreover, WSRT radio
data at $1.4-5$ GHz indicate that there is probably an even wider
outflow of yet slower material (Rol \etal\ in preparation).  It is not
clear however, that this material produces enough X-ray emission to
explain the slow late decays, and so the cause of these may be
altogether different.

From Figure 3 alone, it is not clear whether ordinary GRBs and \grb\
are two distinct classes of events with different X-ray light curves,
or they form the boundaries of a continuum of high energy transients
that will eventually fill in the entire left side of the figure. A
simple theoretical interpretation discussed in \Sect{theory} favors a
continuum of events. As we discuss, at early times an off-axis observer sees a
rising light curve, peaking when the jet Lorentz factor
is $\sim 1/\theta_{\rm obs}$, and approaching that seen by an on-axis
observer at late times. This leads us to believe that the low X-ray
luminosity of the recently discovered GRB~031203 may have also been due to its having being viewed substantially off-axis.

\subsection{The X-Ray Light Curve of Sources S1 and S2}
\lSect{xlite}

Figure 4 summarizes the X-ray observations of sources S1 and S2. The
first four data points (or upper limits) up to day 200 are from \sax\
(Pian \etal\ 2000) and do not resolve individual sources within S1 and
S2; our \cxo\ observations on day 1281 do resolve the sources. We
consider two hypotheses: 1) that \sn\ and \grb\ were the same event,
both happening within S1, and 2) that \grb\ occurred within the error
box of S2 at a cosmological distance and was thus a more ordinary
GRB. Hence the S2 observations are plotted at distance, $z = 1.34$,
such that the afterglow luminosity on day 1 is comparable to ordinary
GRBs (see also \Sect{systematic}). The S1 observations are plotted
with an assumed distance of 35.6 Mpc, the distance to the
supernova. The subtraction of the known fluxes of sources S1b and S1c
on day 1281 from the four \sax\ points reduces their values by
$\sim4\%$, which is less than the size of the symbols used in the
plot. We chose instead to plot the sum of all three S1 sources in
Figure 4 (S1$_{\rm sum}$) to indicate the ``expected'' flux from
extrapolation of the \sax\ X-ray light curve of S1 and compare it with
the flux of the S1a (the \sn) point only.

In contrast to the curve for S1 or ordinary GRBs (Figure 3), the curve
for S2 shows random variability and is nearly flat (assuming
S(3$\arcmin$) for the \cxo\ flux of S2 on day 1281). It is consistent
with a collection of variable X-ray sources, probably distant AGNs,
whose sum sometimes exceeds and at other times falls below the \sax\
threshold. The total flux of the \cxo\ observations of the sources
within the larger (3$\arcmin$ radius) error circle of S2 is within a
factor of 1.5 of the brightest flux ever detected by \sax\ for S2 (day
1) and the two upper limits given by \sax\ on days 2 and 200. If the
GRB occurred within S2 it either declined more rapidly than any other
GRB ever studied before, in which case the observations of S2 offer no
supporting evidence for the connection, or it created a most unusual
afterglow that has not declined, in over three years. Moreover, the
afterglow on day 1281 would have a luminosity orders of magnitude
greater than other GRBs after day 50 (Figure 3). The simplest
conclusion is that S2 did not contain \grb\ and that the \sax\
detection was a collection of variable background sources.

The light curve of S1, on the other hand, shows a gentle decline to
day 200, followed by a rapid fading, by factor of about 12 to
$1.1\times10^{39}$ erg s$^{-1}$, by day 1281 (here we compare the
\sax\ value at day 1 to S1$_{\rm sum}$).  This last data point is
consistent with the light curves of other particularly luminous
supernovae, e.g., SN~1993J, that may have had high mass loss rates,
but one must take care because SN 1998bw was a Type Ic supernova, which
presumably occurred in a Wolf-Rayet star. Such stars are known to have
a high wind velocity, and hence a low circumstellar density. Bregman
et al. (2003) have recently discussed the X-ray emission of young
supernovae and find that most have an X-ray luminosity in the $0.5-2$
keV energy band less than $2 \times 10^{39}$ erg s$^{-1}$.  They do
point out exceptional cases - SN~1978K, SN~1996J, SN~1998Z, SN~1995N,
and SN~1998S - that have X-ray luminosities from 10$^{39.5}$ to
10$^{41}$ erg s$^{-1}$ even a decade after the event, but these were
all Type II.  Given the exceptionally large kinetic energy inferred
for \sn\ (Iwamoto et al. 1998; Woosley et al. 1999), perhaps it is not
surprising that its luminosity after three years should place, e.g.,
about an order of magnitude above common Type Ic supernovae like SN
1994I.

The simplest conclusion here is that the brilliant emission of S1a
during the first days was the X-ray afterglow of the relativistic
ejecta that made \grb. By day 1281 however, we were seeing the
energetic, but not especially relativistic ejecta of \sn\ colliding
with the presupernova mass loss of its progenitor star. This
hypothesis is discussed further in \Sect{theory}.

\section{Theoretical Interpretation}
\lSect{theory}

The X-ray afterglow emission of ordinary GRBs is generally attributed
to synchrotron emission from shocks as the blast encounters the
interstellar or circumstellar medium. Some useful scaling relations
for blast waves in which each particle emits a fixed fraction,
$\epsilon$, of the energy it gains in the shock, have been given by
Cohen, Piran \& Sari (1998).
\begin{equation}
L(t) \propto t^{-[(m-3)/(m+1)]-1}
\end{equation}
with 
\begin{equation}
m = \frac{\epsilon^2 + 14 \epsilon + 9}{3 - \epsilon}.
\end{equation}
Fundamentally, $ 0 < \epsilon < 1$ and $3 < m < 12$, so that $L
\propto t^{-1}$ to $t^{-22/13} = t^{-1.69}$. This expression is for
constant density. Chevalier \& Li (2000) also give expressions for the
power law scaling of afterglow light curves and find for a medium with
$\rho \propto r^{-2}$,
\begin{equation}
L(t) \propto t^{-\alpha}
\end{equation}
with $\alpha$ 1.75 to 2.17 for radiative blast waves and 1.38 to 1.75
for adiabatic blast waves if the index of the electron power
distribution, $p$, is between 2.5 and 3.0.  The curve $L = 10^{46}
t^{-1.69}$ erg s$^{-1}$ is plotted in Figure 3, and provides a
reasonable description of the early X-ray afterglow lightcurve (when
most of the energy is emitted at these frequencies). At observer times
longer than a week, the blast wave would, however, be decelerated to a
moderate Lorentz factor, irrespective of the initial value. The
beaming and aberration effects are thereafter less extreme. If the
outflow is beamed, a decline in the light curve is expected at the time when the inverse of the bulk Lorentz factor equals the opening angle of the outflow (Rhoads 1997). If the critical Lorentz
factor is less than 3 or so (i.e. the opening angle exceeds
20$^\circ$) such a transition might be masked by the transition from
ultrarelativistic to mildly relativistic flow, so quite generically it
would be difficult to limit the late-time afterglow opening angle in this
way if it exceeds 20$^\circ$. For reasonable conditions then, the
power law declines of both XRFs and luminous GRBs, given in Figure 3,
are what might be expected from very relativistic ejecta slowing in
either a constant density medium or a circumstellar wind.

However, the very slow decline of \sn\ needs a different
explanation. \grb, or at least that portion directed at us, was very
weak. The total energy of its relativistic ejecta has been estimated as no
more than $3 \times 10^{50}$ erg (Li \& Chevalier 1999) and most of
that probably was not directed toward us. Wieringa, Kulkarni \& Frail
(1999) discuss the possibility of two shocks associated with \sn, a
relativistic one from the GRB and a slower moving, more powerful shock
from the supernova, but until very late times when the blast reaches
the termination of the presupernova wind, the thermal emission from
the supernova itself is expected to be weak because of the low density
of the wind (Chevalier 2000).

Two possibilities can be considered. First, that \grb\ was an
``ordinary'' (or somewhat subluminous) GRB observed off-axis. This has
been suggested many times (e.g. Woosley, Eastman \& Schmidt 1999;
Nakamura 1999; Granot, Panaitescu, Kumar \& Woosley 2002; Yamazaki,
Yonetoku \& Nakamura 2003; Zhang, Woosley \& Heger 2003) with
different underlying assumptions regarding the angular distribution of
the ejecta. Nakamura (1999) assumes that the jet has sharp edges and the
peripheral emission comes from scattering. Woosley \etal\ (1999) assume that
there is a distribution of ejecta energies and Lorentz factors and
that, during the burst, we see only the low energy wing moving toward
us.  The other possibility is that \grb\ was deficient in energetic
gamma-rays at all angles. This is not incompatible with the fact that
it may have ejected $3 \times 10^{50}$ erg of mildly relativistic
($\Gamma > 2$) material, but concerns only the very relativistic
ejecta, $\Gamma > 200$, thought responsible for harder GRBs.

The X-ray light curve can help to distinguish these two possibilities.
First, energetic as it may be, at early times the X-ray emission of
the underlying subrelativistic supernova is probably
negligible. Though the mass loss rate of the Wolf-Rayet star
progenitor may have been high, the wind velocity was also large
implying a low circumstellar density and inefficient supernova
emission (Chevalier 2000). Type Ib and Ic supernovae are typically
weak X-ray emitters (Figure 3; Bregman 2003). If we attribute the
X-ray emission during the first 200 days to a weak, but on-axis GRB,
then the X-ray afterglow should have faded with a power law not too
different from the other GRBs. The fact that its decay is nearly flat is
inconsistent with {\sl any} relativistic blast wave in which the
electrons emit a constant fraction of the energy gained in the shock
(Cohen, Piran \& Sari 1998), even in a constant density medium, and argues
against an explosion with a single energy and Lorentz factor seen pole
on.

The data may be more consistent with a powerful burst seen off-axis
(Granot et al.  2002). Even along its axis, the burst may have been
weaker than most, but the energy per solid angle could still have been
orders of magnitude greater than along our line of sight, accounting
for most of the $\sim 3 \times 10^{50}$ erg inferred from the radio
(Li \& Chevalier 1999). Because of relativistic beaming, initially we
see only the low energy material moving toward us, but as the core of
the jet decelerates, its afterglow is beamed to an increasing angle so
that more and more energetic material becomes visible along our line
of sight. Depending on the geometry, the afterglow luminosity could
even temporarily increase. Granot et al. (2002) consider the
appearance of \grb\ at various angles and conclude that the viewing
angle needs to have been $\gtaprx 3 \theta_o$ with $\theta_o$ the half
angle of the most energetic part of the jet, otherwise the optical
afterglow would have contaminated the supernova light curve
unacceptably. One expects that the X-ray light curve would look
similar to some of the plots of Granot et al. with the critical
addition of low energy wings of ejecta as calculated by Zhang, Woosley
\& Heger (2003). This material would raise the luminosity at early
times when almost nothing is seen of the central jet.  A calculation
of this sort must be a high priority for the community.
 
As time passes, beaming becomes less important and the entire
decelerating jet becomes visible, followed a little later by the
underlying supernova. The light curve should then decline, as it did
between days 200 and 1281 in Figure 3. One can estimate the time scale
for when beaming becomes unimportant.  For a circumstellar density
distribution $\rho = A r^{-2}$ Waxman, Kulkarni \& Frail (1998),
assuming typical mass loss and GRB parameters, estimate a radius
\begin{equation}
r_{\rm NR} \ = \ \frac{E}{4 \pi A c^2} \ = \ 1.8 \times 10^{18}
\frac{E_{52}}{A_*} \ {\rm cm},
\end{equation}
and a corresponding time
\begin{equation}
t_{\rm NR} \ = \ \frac{r_{\rm NR}}{c} \ = \ 1.9 \, (1 + z)
\frac{E_{52}}{A_*} \ {\rm yr}
\end{equation}
when the explosion has swept up a rest mass comparable to the initial
relativistic ejecta. Here, $A_* = A/5 \times 10^{11}$ g cm$^{-1}$,
corresponding to a fiducial mass loss rate of $1.0 \times 10^{-5}$
M$_\sun$ \ yr$^{-1}$ for a wind velocity of $1000$ km/s, and E$_{52}$
is the relativistic energy in $10^{52}$ erg s$^{-1}$. One thus expects
that for times of order several years the relativistic energy will
have been radiated away and the emission will become isotropic. We can
define this as the onset of the supernova stage.

Evidence of a close association of the early X-ray emission to the
overall GRB phenomenon is also presented by Berger, 
Kulkarni \& Frail (2003). They have studied a sample of 41 GRB X-ray 
afterglows and they find a strong correlation between their X-ray
isotropic 
luminosities ($L_{\rm X,iso}$) (normalized to $t=10$ hours after
trigger) and their beaming fractions. We plot 
$L_{\rm X,iso}$ as a function of the GRB isotropic 
equivalent $\gamma-$ray energy, $E_{\rm iso}$ (Figure 5). The 
data are taken from Berger, Kulkarni \& Frail (2003) and we
 have added data for GRBs 031203, 030329 and 980425, as indicated on
the plot. Several points are striking in this empirical relation: GRBs 
031203 and 980425 fall well within the overall correlation and 
{\it extend} the association by roughly three orders of magnitude. 
We have fitted the data without (dashed line) and with (solid line)
 all the outliers (980425, 031203,000210, 990705), with a power law 
of index 0.61 and 0.72, respectively; we conclude that the data are
 consistent with a trend extending roughly six orders of magnitude 
in X-ray luminosity. Parenthetically, the two outliers in the plot 
correspond to GRB~000210 (a `dark' GRB) and GRB~990705 (a very 
bright GRB), and indicate that it may be possible to distinguish 
GRB subclasses by simply using their X and $\gamma-$ray 
properties, as also pointed out by Berger, Kulkarni \& Frail (2003). 

Note that the convergence to a ``supernova'' at three to ten years here does not
require all XRFs and GRBs to have a bright optical supernova following
the GRB. Bright optical emission is a statement about the
radioactivity the supernova made. X-ray emission at 3 years is about
its kinetic energy.  If the variation in GRB energy and X-ray light
curve is simply an effect of viewing angle, then one expects the empty
parameter space in Figures 3 and 5 to fill in with future observations. The
recent GRB~031203 (Figures 3, 5) may well be the first of these `gap'
events (Rodriguez-Pascual \etal\ 2003; Soderberg \etal\ 2003a) if the
reported redshift of $z=0.105$ (Prochaska \etal\ 2003) is confirmed.

\section{Conclusions}

Our study resulted in the detection, on day 1281, of multiple X-ray
point sources in the two \sax\ error boxes S1 and S2 originally given
(Pian \etal\ 2000) for the variable X-ray counterpart to \grb\
(\Sect{sources}).  Based upon the accurately known radio location, the
source S1a is definitely the supernova. The sum of all sources in S2
on day 1281 is similar to the maximum observed for this error box by
\sax\ by a factor that is smaller than 1.5. This is consistent with
the hypothesis that S2 did not contain the GRB, but instead some
variable X-ray sources that have not disappeared. S1, however, has
been consistently observed to decline for $\sim1300$ days. We conclude
that the source S1 contained the GRB and that \sn\ and \grb\ were the
same explosion.

Additional insight comes when we compare the X-ray light curve of
\grb\ with the broader family of high energy transients
(\Sect{systematic}). Comparison of published X-ray light curves for
GRB, XRF and SNR supports a {\it unification hypothesis}, similar to
that proposed for AGN. Under this hypothesis all of these sources are
associated with standard supernova explosions of massive stars. We can
also distinguish a {\it strong unification hypothesis} in which the
properties of the external medium are also standardized. There is
considerable, though still inconclusive, evidence that the strong
unification hypothesis is false, though it is still worth testing.
These explosions are conjectured to produce an anisotropic, beamed
component associated with a decelerating, ultrarelativistic outflow
and an unbeamed, isotropic component associated with the slowly
expanding stellar debris. The flux associated with the beamed
component depends upon the observer direction and declines rapidly
with observer time; the closer to the axis the larger the flux and the
more rapid is the onset to the typical afterglow decline. As the
beamed component decelerates to become entirely non-relativistic, the
observed flux will become independent of orientation. Under the strong
unification hypothesis a one-parameter (inclination) family of X-ray
light curves will be produced, converging asymptotically to a single
variation when the beamed component becomes non-relativistic. If only
the unification hypothesis is true, we should still be able to observe
these trends and estimate the inclination.  Indeed, this is roughly
what we observe when we plot the isotropic luminosities of GRB, XRF
and SNR on a common scale (Fig.~3).  It appears that after three years
all explosions are subrelativistic, with X-ray luminosity dominated by
the stellar debris, $\sim10^{39}$~erg s$^{-1}$. We therefore
tentatively identify GRB, XRF and SNR as similar objects observed with
small, medium and large inclination respectively.  More specific to
this paper, the observation that \sn\ and \grb\ follow a smooth light
curve, which fits this pattern supports the claim that they are the
same source.

In \Sect{theory} we discussed two possible interpretations of these
light curves based upon either a standard phenomenon viewed at
different angles or explosions that eject variable amounts of
relativistic ejecta. We conclude that the observations, especially the
slow initial decline rate, are more consistent with the ``off-axis
model'' in which \grb\ was a much more powerful GRB seen at an angle
greater than about three times the opening angle of the central jet.
Emission at early times does not come from this central jet but from
wings of less energetic material. After about three years the emission
of all these high energy transients becomes isotropic and we see the
sub-relativistic ejecta of the supernova interacting with the
circumstellar wind. Thus all high energy transients have a common
luminosity at three years due to their non-relativistic ejecta, but
they follow different decay rates, depending upon the viewing angle,
to reach there.

Further, we discussed the stellar environment in that region of the
galaxy ESO~$184-$G$82$ where the supernova occurred. The supernova is
one member of an X-ray doublet, both of which seem to be in the galaxy
and to have spend part of their life cycles as ULXs.  The projected
distance between the two X-ray sources is $\sim300$ pc, which is
suggestive of some sort of a very active star forming region.

Finally, we would like to stress that the existence of a relation between the decline rate of the X-ray light curve during the first few weeks and its brightness implies that
such measurements might be useful for diagnosing the character and
subsequent evolution of a given high energy transient as well as
constraining its distance. However, the sparse coverage of the current X-ray afterglow data does not allow us to address fundamental questions such as: Did the rapid decline of \grb\ continue?  Did it/will it level out above or in the vicinity of other Type Ib/c
supernovae?  At what point does \sn\ become like ordinary supernovae? 
The current data enable us to make a {\it prediction}
on the unification of the GRB-SN phenomena, which can only be
vindicated with further observations, obviously of \sn, but also of
the nearest XRFs and GRBs to fill in the missing parameter space. We strongly encourage, therefore, followup observations of nearby GRBs and XRFs for as long as the available instrumentation allows; we also encourage the calculation of off-axis models of the
X-ray light curves, especially for a variable distribution of Lorentz factors and energies.

\acknowledgements {We are grateful to Re'em Sari for educational
conversations regarding the nature of X-ray afterglows. CK and SP
acknowledge support from SAO grant GO1-2055X. SW is supported by the
NASA Theory Program (NAG5-12036), and the DOE Program for Scientific
Discovery through Advanced Computing (SciDAC; DE-FC02-01ER41176)}. ERR
has been supported by NASA through a Chandra Postdoctoral Fellowship
award PF3-40028. Part of this work was performed while CK, ERR and RB
were visiting the University of CA in Santa Cruz. The authors
acknowledge support (RAMJW) and benefits from collaboration within the
Research Training Network ``Gamma Ray Bursts: An Enigma and a Tool'',
funded by the EU under contract HPRN-CT-2002-00294.

\begin{figure}
\label{fig1}
\figurenum{1}
\begin{center}
\epsscale{0.78}
\vspace{-1.6truecm}
\plotone{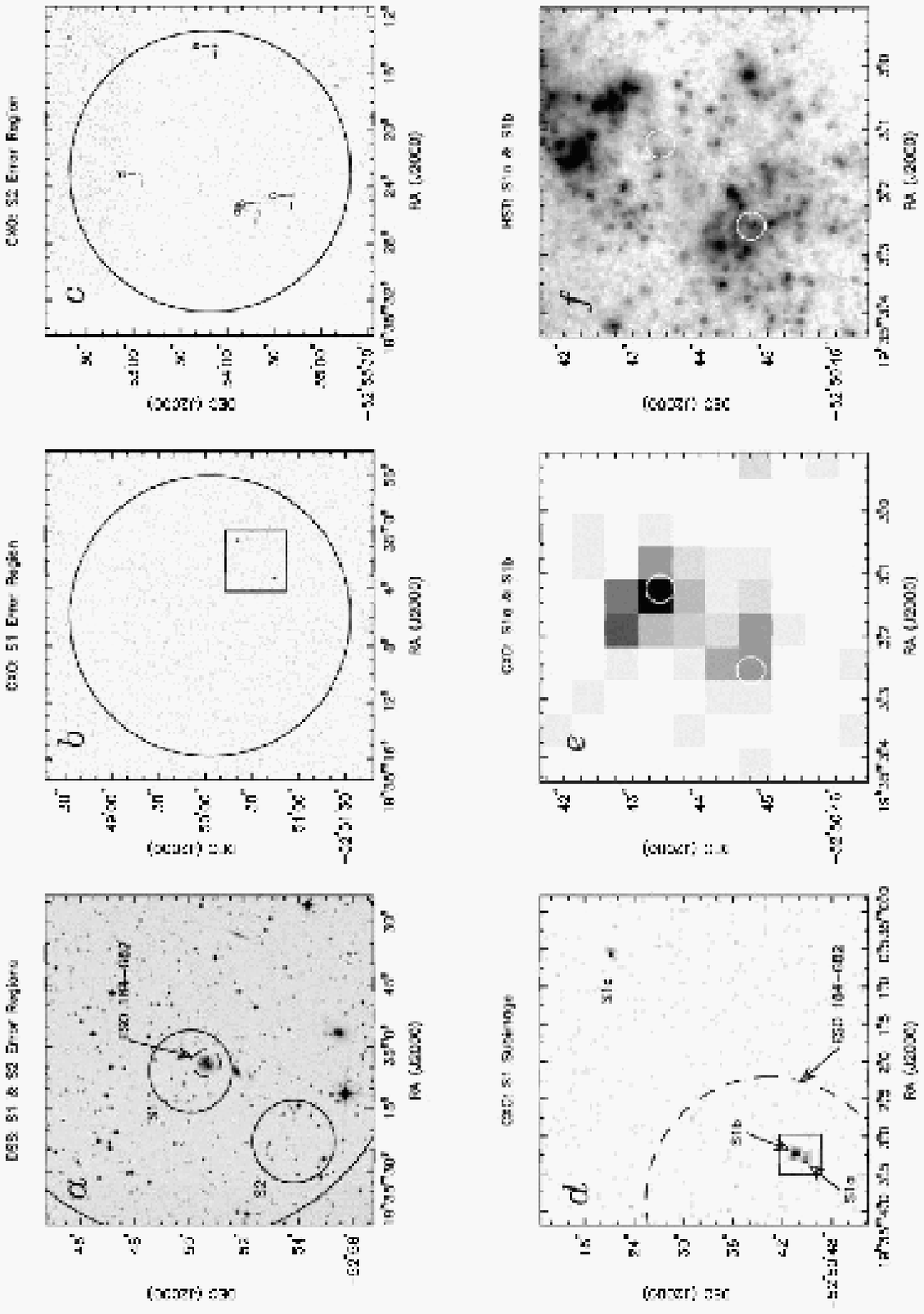}
\end{center}
\vspace{-1.1truecm}
\caption{
({\sl a}): Digitized Sky Survey (DSS) image of the area around \grb. The outer (partial) circle represents the 8.0\arcmin\ error region for \grb\ from the \sax\ {\it WFC}.
The full solid circles represent the 1.5\arcmin\ \sax\ \nfi\ S1 and S2
error regions.  The dashed circle (radius$=15\arcsec$) within S1
identifies the area covered by the \grb\ host galaxy ESO~$182-G82$.
({\sl b}):\cxo\ image of S1. The boxed region containing S1ab
(lower left) and S1c (upper right) is shown in ({\sl d}).
({\sl c}):\cxo\ image of S2. Almost the entire error region (over
97\%) was covered by the \cxo\ CCDs S2 and S3. Source S2d falls into the gap between these two CCDs. ({\sl d}): \cxo\ blow up of
the square box in ({\sl b}).  The dashed circle (radius$=15\arcsec$)
indicates the host galaxy area (also identified in ({\sl a})).  A
close-up view of the solid box shown here is seen in ({\sl e}).
({\sl e}):\cxo\ blow up of the square box in ({\sl d}). Sources S1a
and S1b are clearly resolved. S1a is the source to the south and
coincides with the established radio location of \sn. The two solid
circles represent the 1$\sigma$ error regions for these sources with
($r=0.3\arcsec$) derived with comparative astrometry.
({\sl f}):{\it HST/STIS} image of ({\sl e}) taken in the 50CCD/Clear mode (no filter). The
white circles are the \cxo\ error circles from ({\sl e}).}
\end{figure}

\begin{figure}
\label{fig2}
\figurenum{2}
\begin{center}
\epsscale{0.75}
\rotatebox{-90}{\plotone{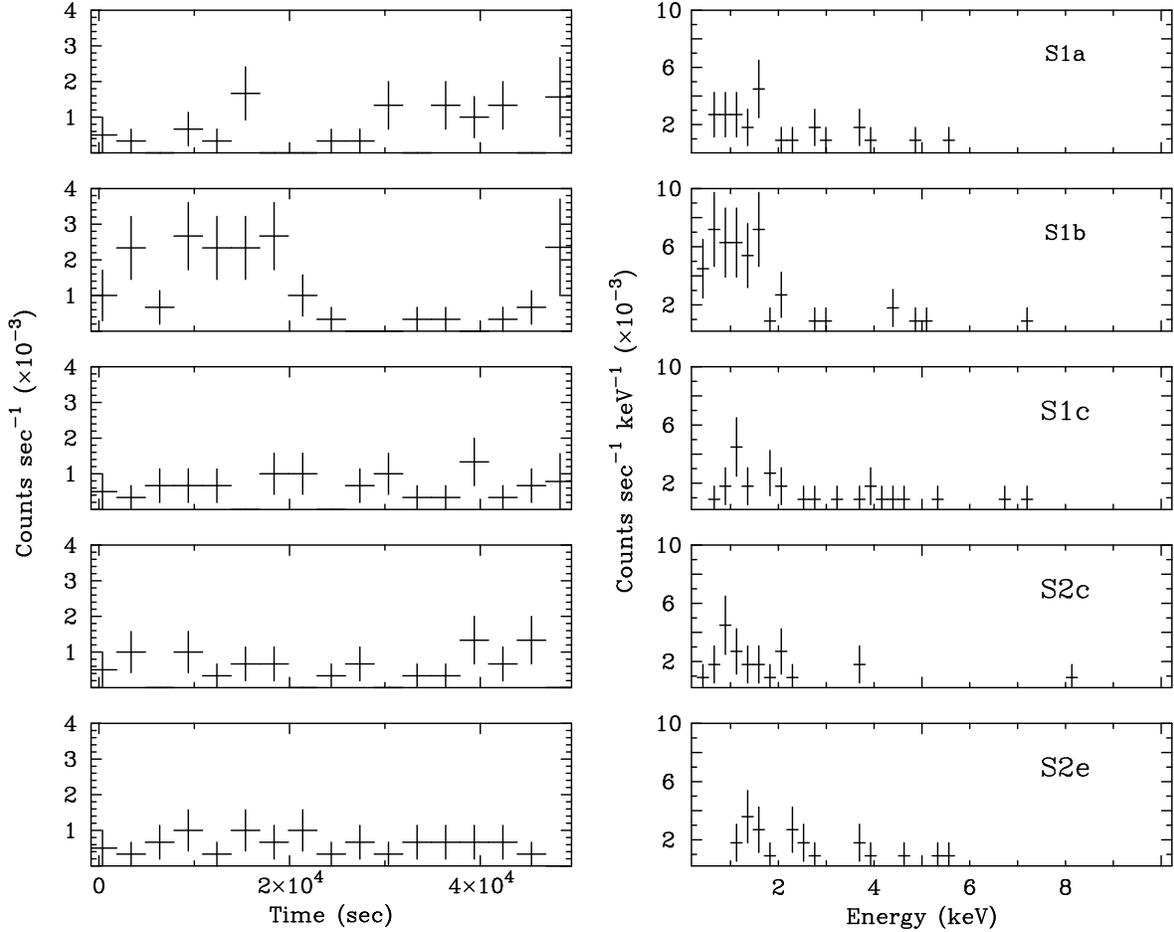}}
\end{center}
\caption{Light curves (left column; binned in 3000 s wide bins) and spectra (right column; binned in 0.232 keV wide bins) of the five sources with more than 20 counts total within S1 and S2 collected with \cxo\ ($0.3-10.0$ keV). Notice source S1b is highly variable during the 50 ks observation. All other sources are either too faint to determine any variability or consistent with a constant persistent emission.}
\end{figure}

\clearpage

\begin{figure}
\label{fig3}
\figurenum{3}
\begin{center}
\epsscale{0.70}
\rotatebox{-90}{\plotone{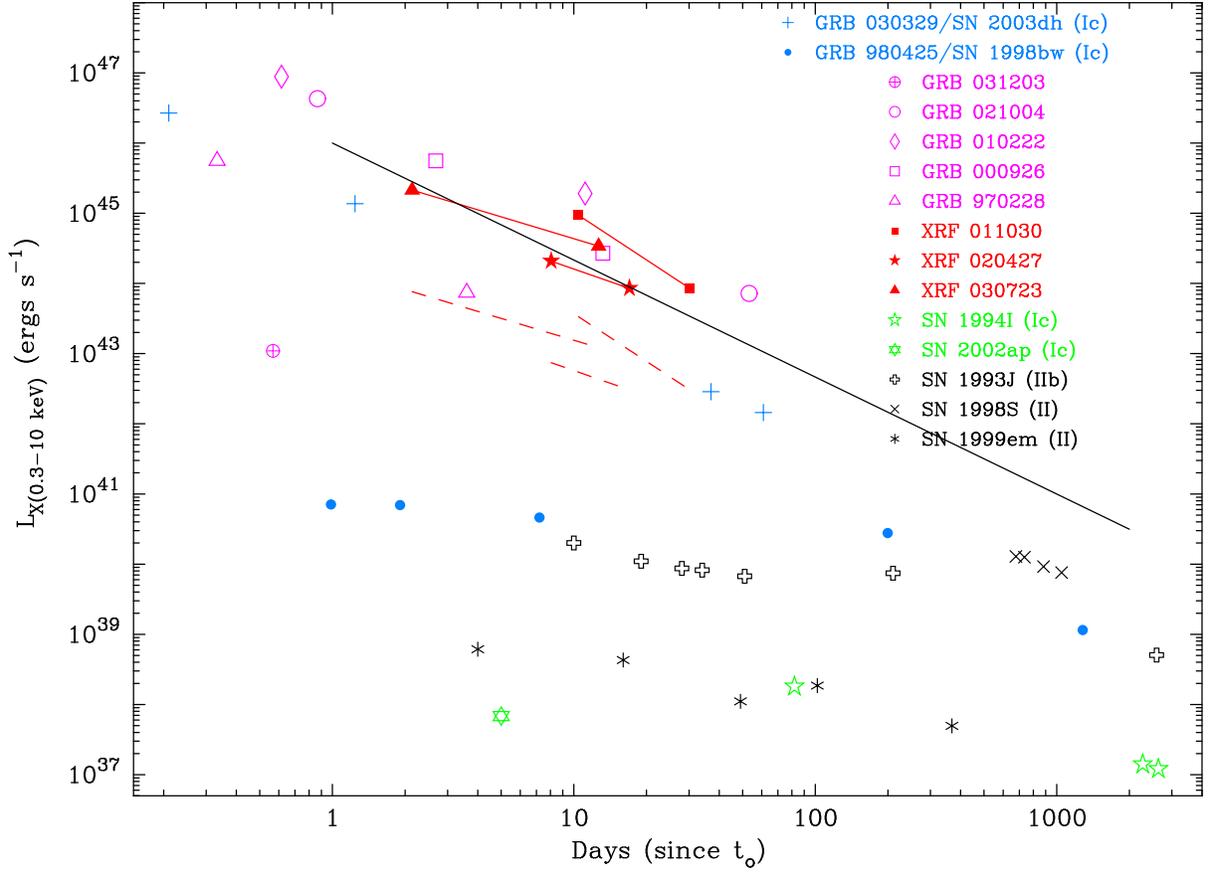}}
\end{center}
\caption{Compilation of GRB, XRF, SN I and SN II X-ray light curves
($0.3-10.0$ keV) presented as (isotropic) luminosity distances using a
cosmology with $\Omega_{\rm M}=0.27, \Omega_{\Lambda}=0.73$, and
$H_0=72$ km s${-1}$ Mpc${-1}$. The XRF luminosities are calculated
assuming two redshifts, $z=1$ (solid lines) and $z=0.251$ (dashed lines). The solid long line corresponds to a temporal decay of $10^{46}$ t$_{\rm days}^{(-1.69)}$, discussed in the text.}
\end{figure}

\begin{figure}
\label{fig4}
\figurenum{4}
\begin{center}
\epsscale{0.7}
%\rotatebox{-90}{\plotone{grb_sn_f4.ps}}
\rotatebox{-90}{\plotone{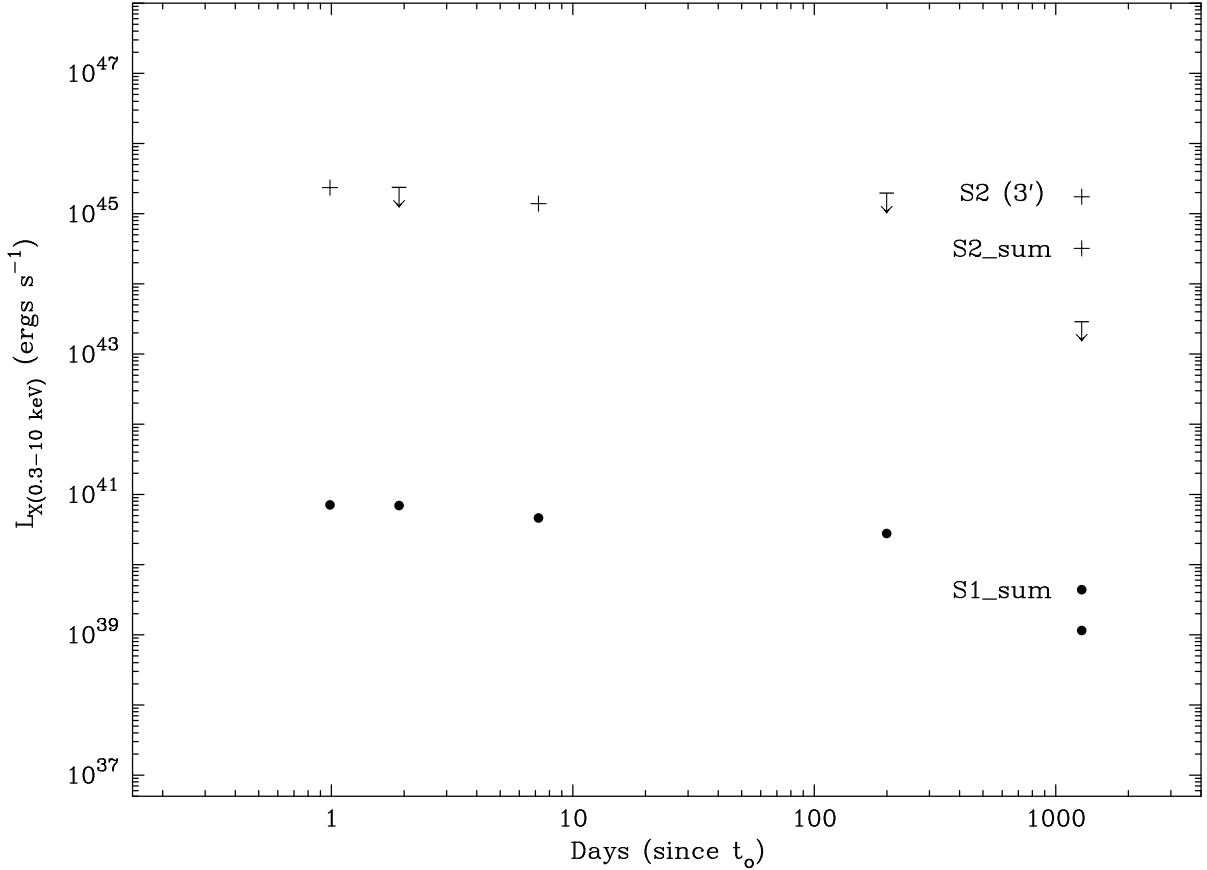}}
\end{center}
\caption{The X-ray light curves of sources S1 and S2. The upper curve
gives the two \sax\ observations and upper limits for
the error box of S2 during the first 200 days after the GRB trigger. The last
points (day 1281) on that curve are the sum (S2$_{\rm sum}$) of all point sources observed with \cxo\ and the sum (S2(3$\arcmin$)) of all sources observed with \cxo\ within an error circle of 3$\arcmin$ radius. A distance of $z=1.343$ has been
assumed for all S2 points (see text). The arrow on day 1281 reflects the \cxo\ detection limit of a source placed at $z=1.343$ or less.
The lower curve is the light curve of source S1 assuming a
distance equal to that of \host\ (35.6 Mpc). The first four points
are the \sax\ observations; the fifth (upper) point at day
1281 is the sum of all S1 \cxo\ sources (S1$_{\rm sum}$). The point below S1$_{\rm sum}$ is the \sn\ luminosity after we subtract the flux contribution from sources S1b and S1c from S1$_{\rm sum}$.}
\end{figure}

\begin{figure}
\label{fig5}
\figurenum{5}
\begin{center}
\epsscale{0.95}
\plotone{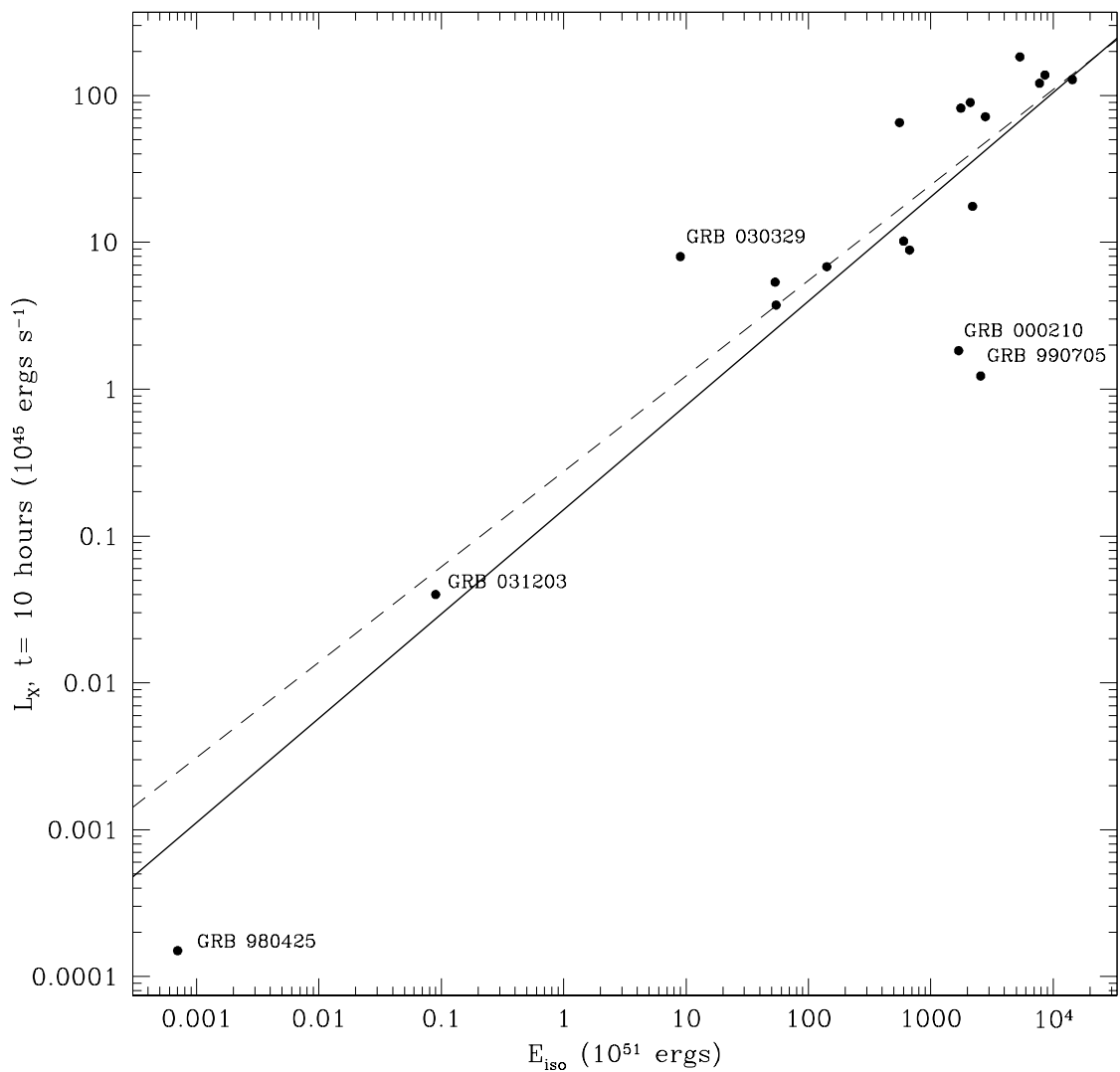}
\end{center}
\caption{Isotropic equivalent luminosity of GRB X-ray afterglows, $L_{\rm X}$, scaled to $t=10$ hours after the burst as a function of their isotropic equivalent $\gamma-$ray energy, $E_{\rm iso}$. The data are taken from Berger, Kulkarni \& Frail (2003); here we have added data for GRBs 031203, 030329 and 980425, as indicated on the plot. The solid line is a fit including all data; the dashed line is a fit exluding GRBs 980425, 031203, 000210, and 990705.}
\end{figure}

\end{document}